\documentclass[pra,a4paper,twocolumn,showpacs,amsmath,amssymb,amsfonts]{revtex4}
\usepackage{bm}
\begin{document}

\title{RESULTS OF THE INVESTIGATIONS OF THE NATURE OF THE LONG
-TERM INSTABILITY IN QUANTUM FREQUENCY STANDARDS AND
MAGNETOMETERS}

\author{A.G.Chirkov}

\email{agc@ac11593.spb.edu}

\affiliation{Department of Theoretical Physics, State Polytechnic
University, St.-Petersburg, 195251, Russia}

\date{\today}

\begin{abstract}
The problem analysis results made the author to draw a conclusion
that the nature of the resonance frequency long-term instability
and drift at harmonic excitation is related to the phase dynamics
of the "atom + field" system in the small $\varepsilon $ -
vicinity of the resonance. The investigation is based on the
strictly substantiated asymptotic Krylov-Bogolyubov perturbation
theory. A time-dependent (drift) first-order correction $\delta
\omega ^{\left( 1 \right)}$ of the perturbing field amplitude $E_1
\left( {H_1 } \right)$\textsc{ }to the resonance frequency $\omega
_0 $ was disclosed. It was found that this correction is always
present and is responsible for the frequency drift and long-term
instability. The necessary and sufficient conditions of accurate
resonance, as well as the conditions of realization of a stable
(stationary, steady-state) drift-free oscillation regime in a
quantum system, are obtained.
\end{abstract}

.

\maketitle

\section{Introductory remarks}

Intensive investigations carried out by different research centers
with the aim of development of new physical principles of
designing frequency (time) standards of different applications
resulted in the high absolute accuracy and short-term instability
of frequency $10^{ - 14}\tau ^{ - 1 \mathord{\left/ {\vphantom {1
2}} \right. \kern-\nulldelimiterspace} 2}$ over a time averaging
period $< 10^4$ s. It is well known that the long-term instability
of precision and reference atomic frequency standards over a time
period of a day is an order of magnitude worse than the short-term
instability. The difference in these parameters exists also in new
devices, such as ion trap and atomic fountain and in active and
passive hydrogen frequency standards. The attainment of a
long-term instability comparable with the short-term instability
remains a problem, i.e. frequency drift remains nonremovable, its
sign and value are unpredictable, and its nature is still unknown.

Some authors tried to attribute the atomic standard, frequency
drifts, perceptible even within over time periods $\tau > 10^4$s,
to possible variability of the physical constants - the gravity
constant G \cite{1} and the fine structure constant $\alpha $
\cite{2}. Though, another author \cite{3}, based on the
experimental data analysis, adduces arguments showing the
groundlessness of this approach to explain the standard frequency
drifts, at least, within time periods under consideration - a
month, ...a year, which are extremely short with respect to
cosmological measures.

In evaluating standard frequency stability, the "two-selectivity
variation" model, proposed by D.Allan, is preferred. This model
eliminates the slow drift. Such an evaluation gives a more
pleasing result, but doesn't show the true state of affairs.
Theoretical investigations of the interaction of a two-level
system with a harmonic field for the special case of the weak
field ${E_1 } \mathord{\left/ {\vphantom {{E_1 } {E_0 ({H_1 }
\mathord{\left/ {\vphantom {{H_1 } {H_0 ) \ll 1}}} \right.
\kern-\nulldelimiterspace} {H_0 ) \ll 1}}}} \right.
\kern-\nulldelimiterspace} {E_0 ({H_1 } \mathord{\left/ {\vphantom
{{H_1 } {H_0 ) \ll 1}}} \right. \kern-\nulldelimiterspace} {H_0 )
\ll 1}}$, are described in many papers. This situation is realized
in high-accuracy quantum devices under consideration. The papers
point out one important result the resonance occurs at the
frequency $\omega = \omega _0 + \delta \omega ^{\left( 2 \right)}$
which differs from the frequency of the unperturbed
transition$\omega _0 $. The arising constant correction $\delta
\omega ^{(2)} = {(1} \mathord{\left/ {\vphantom {{(1} {4)\gamma
}}} \right. \kern-\nulldelimiterspace} {4)\gamma }H_1^2 / H_0 $,
is of the second order of smallness of the disturbing field $H_1 $
(or $E_1 $,) amplitude, and is known as the Bloch-Siegert shift.
The value of the correction is smaller than $10^{ - 6}\% $ and in
practice it is neglected in most cases.

However, in spite of the progress made towards the understanding
of the two-level system dynamics, particularly, of the resonance
phenomena in spin-systems, not a single of the theoretical papers,
known to us, provides any information on the existence of the slow
drift of the resonance center frequency in quantum systems.

\subsection{The problem analysis results}

The analysis of the problem from different view points has led us
to a number of conclusions.

The generally accepted condition for the resonance $\Delta \omega
= \left| {\omega - \omega _0 } \right| = 0$ is \textit{incomplete.
}On theory and in practice, it is customary to assume that the
condition for the exact resonance is the equality of the
difference in tuning frequencies $\Delta \omega $ to zero. In
practice, this condition is tried to be fulfilled with the highest
accuracy through the use of up-to-date facilities of computerized
tracking of the resonance center. However, when the resonance
phase-frequency characteristic $\Delta \varphi $ is taken into
account, the equality of detuning $\Delta \omega = \left| {\omega
- \omega _0 } \right| = 0$ only allows for the first condition of
resonance (coherence). On the second resonance condition the
theory shows that at zero detuning the oscillation phase
difference $\Delta \varphi $ between the field $\vec {H}_1
$\textsc{ }and the atom should be equal to$( - \pi \mathord{\left/
{\vphantom {\pi {2)}}} \right. \kern-\nulldelimiterspace} {2)}$.
The investigations that rigorously prove practical fulfillment of
the second conditions of coherence in a quantum system arc
lacking. The verification of practicability of this condition
calls for investigations of the phase state dynamics of the
"atom+field" system in the infinitesimal vicinity of resonance.

The basis for the resonance frequency drift is not a technical
reason but an obscure physical effect, which brings about the
instability of resonance regime of oscillations, and the
instability cannot be obviated with a technical means.

\subsection{\label{sec:level2}Statement and solution of the problem}

This paper shows the solutions of the following problems
\cite{4,5}:

1). Investigation of a phase state dynamics of the two-level
system "atom + field" in a weak variable field in a small vicinity
of a resonance.

2). The necessary and sufficient conditions for the exact
resonance.

3). Existence of the stationary resonant oscillation regime and
its steadiness.

The research was carried out using the strictly substantiated
asymptotic Krylov-Bogolyubov theory of perturbations \cite{6,7}
which allows studying the system over any time intervals $t \sim 1
\mathord{\left/ {\vphantom {1 \varepsilon }} \right.
\kern-\nulldelimiterspace} \varepsilon $ ($0 < \varepsilon \ll
1)$.

The results of the researches allow to explain the nature of
long-term frequency (time) instability in reference quantum
devices.

\section{EQUATION OF THE EXACT FREQUENCY OF A PERTURBED QUANTUM SYSTEM}

\subsection{Equation with a small parameter}

The typical equations for a density matrix of two-level system
interacting with an external weak variable magnetic or electrical
field are \cite{8,9}

\begin{equation}
\label{eq1}
\begin{array}{l}
 \dot {\rho }_{22} = \Lambda _2 - \Gamma _2 \rho _{22} - iV\left( {\rho
_{21} - \rho _{12} } \right) \\
 \dot {\rho }_{11} = \Lambda _1 - \Gamma _1 \rho _{11} + \;\;iV\left( {\rho
_{21} - \rho _{12} } \right) \\
 \dot {\rho }_{21} = - \left( {\Gamma _{21} + i\omega _0 } \right)\rho _{21}
- iV\left( {\rho _{22} - \rho _{11} } \right) \\
 \end{array}
\end{equation}

\noindent$\Lambda _1 ,\,\Lambda _2 $ - rates of non-coherent
pumping on the appropriate level; $\Gamma _1 ,\,\,\Gamma _2
,\,\,\Gamma _{21} $ - relaxation rates; \[ V = \left\langle
{1\left|
{\overset{\lower0.5em\hbox{$\smash{\scriptscriptstyle\frown}$}}{\bm{\mu}
}
\,{\mathbf{\overset{\lower0.5em\hbox{$\smash{\scriptscriptstyle\frown}$}}{V}
}}_{\mathbf{1}} } \right|2} \right\rangle /\hbar
\]- matrix element of interaction;
$\overset{\lower0.5em\hbox{$\smash{\scriptscriptstyle\frown}$}}{\bm{\mu}
} $ - operator of the dipole moment;
${\mathbf{\overset{\lower0.5em\hbox{$\smash{\scriptscriptstyle\frown}$}}{V}
}}_1 \left( t \right) =
{\mathbf{\overset{\lower0.5em\hbox{$\smash{\scriptscriptstyle\frown}$}}{V}
}}_1 \upsilon \left( t \right)$ - intensity of a variable magnetic
(or electrical) field in dipole approach. Let's enter
dimensionless time $t \to \omega \,t$ , the Bloch variables
$R_{1\,} ,\,\,R_{2\,} ,\,\,R_3 $

\[
\begin{array}{l}
 R_1 = \rho _{12} + \rho _{21} \\
 R_2 = - i(\rho _{12} - \rho _{21} ) \\
 R_3 = \rho _{22} - \rho _{11} \\
 \end{array}
\]

\noindent and, differentiating system (\ref{eq1}) with respect to
dimension-less time, we shall write down as

\begin{equation}
\label{eq2}
\begin{array}{l}
 \dot {R}_1 = - \gamma _1 \,R_1 - \nu R_2 \\
 \dot {R}_2 = - \gamma _1 \,R_2 + \nu R_1 + 2\omega _1 \,\upsilon \left( t
\right)R_3 \\
 \dot {R}_3 = \lambda - \gamma _2 \,R_3 - 2\omega _1 \,\upsilon \left( t
\right)R_2 \\
 \end{array}
\end{equation}

The designations normalized to $[\omega ]$ are used in the system
(\ref{eq2}): $\nu = \omega _0 / \left[ \omega \right], \quad
\gamma _1 = \Gamma _1 / \left[ \omega \right],\,\,\,\,\gamma _2 =
\Gamma _2 / \left[ \omega \right]\,\,\,\,,$ $\lambda = \left(
{\Lambda _2 - \Lambda _1 } \right) / \left[ \omega \right]$,
$\omega _1 = \left\langle {1\left|
{\mathord{\buildrel{\lower3pt\hbox{$\scriptscriptstyle\frown$}}\over
{\mu }} \,{\rm {\bf
\mathord{\buildrel{\lower3pt\hbox{$\scriptscriptstyle\frown$}}\over
{V}} }}_1 } \right|2} \right\rangle / \hbar \left[ \omega
\right]$. The basic parameter of the problem $\omega _1 \ll 1$
(approach of a "weak" field). In radio spectroscopy $\omega _1
\sim 10^{ - 4}$, in optics $\omega _1 \sim 10^{ - 8}$. Parameters
$\gamma _1 ,\,\,\gamma _2 ,\,\,\lambda $ are considered as small
as $\omega _1 $. Let's specify it obviously by using the system
(\ref{eq2}) a formal small parameter. Then disturbed system will
be written as

\begin{equation}
\label{eq3}
\begin{array}{l}
 \dot {R}_1 = - \varepsilon \,\gamma _1 \,R_1 - \nu R_2 \\
 \dot {R}_2 = - \varepsilon \,\gamma _1 \,R_2 + \nu R_1 + 2\varepsilon
\,\omega _1 \,\upsilon \left( t \right)R_3 \\
 \dot {R}_3 = \varepsilon \,\lambda - \varepsilon \,\gamma _2 \,R_3 -
2\varepsilon \,\omega _1 \,\upsilon \left( t \right)R_2 \\
 \end{array}
\end{equation}

For investigations of systems (\ref{eq3}) the "action-angle"
variables are usually used \cite{7}. However, instead of the Bloch
variables$R_{1\,} ,\,\,R_{2\,} ,\,\,R_3 $ we shall use the new
variables actually observed $a,\,\psi ,\,z$: $a = \left( {R_1^2 +
R_2^2 } \right)^{1 / 2}$- amplitude of oscillations, $z = R_3$  -
difference of population, $\psi = arctg\left( {R_2 / R_1 }
\right)$ - current phase.

In calculating derivatives for new variables, we shall obtain the
system

\begin{equation}
\label{eq4}
\begin{array}{l}
 \dot {\alpha } = - \varepsilon \,\gamma _1 \,\alpha + 2\varepsilon \,\omega
_1 \,z\,\upsilon \left( t \right)\sin \psi \\
 \dot {z} = \varepsilon \,\lambda - \varepsilon \,\gamma _2 z - 2\varepsilon
\,\omega _1 \,\alpha \,\upsilon \left( t \right)\sin \psi \\
 \dot {\psi } = \,\omega \left( {t,\,\varepsilon } \right) = \nu +
\varepsilon \left( {2\omega _1 \,z / \alpha } \right)\upsilon
\left( t
\right)\cos \psi \\
 \end{array}
\end{equation}

\subsection{The exact resonant frequency}

The last equation is the exact frequency in the perturbed system.
First term is the transition eigenfrequency in the unperturbed
system; shows that the oscillations occur with a constant
frequency $\omega \,(t,\,\varepsilon = 0) = \nu $.

Second term is the generalized correction for a change of the
eigenfrequency under influence of a variable field. This
component, as will be shown below, is responsible for occurrence
of the time constant second order corrections (known as the
Bloch-Siegert shift), third, fourth, orders and the time variable
correction of the first order of the field.

\section{RESONANCE IN TWO-LEVEL SYSTEM}

The perturbing field is a periodic field $\upsilon \left( t
\right) = \cos \omega \,t = \cos t$, where the last $t$ is a
dimensionless value and $[\omega ] = \omega $. The system
(\ref{eq4}) contains the equations with two fast phases: one phase
(nonisochronal) is $\psi $, a role of the second phase
(isochronal) is carried out by $t$. The general research methods
of type (\ref{eq4}) systems are developed in \cite{6,10}. The case
of the main resonance with corresponding equality $1 - \nu = 0$ is
most important.
The experience shows that even a very accurate
equality of the frequencies, $\omega = \omega _0 $ (i.e.$\Delta
\omega = 0)$, there is a slow drift of the centre of the resonance
which is not eliminated\underline { }technically. This makes it
necessary to study the second coherence condition, i.e. the system
phase state in the vicinity of resonance.

\subsection{Dynamics of the system phase state in the $\varepsilon
$ - vicinity of resonance}

For studying the phase state dynamics of the system (\ref{eq4}) in
the å vicinity of the resonance we shall use a new variable
$\vartheta = t - \psi $, which is a difference between oscillation
phases of the field and the atom and is a slow variable in the
$\varepsilon $ - vicinity of resonance. We shall obtain the
equations in the standard Krylov-Bogolyubov form \cite{1,10}

\begin{equation}
\label{eq5}
\begin{array}{l}
 \dot {\alpha } = - \varepsilon \,\gamma _1 \,\alpha - \varepsilon \,\omega
_1 \,z\sin \vartheta + \varepsilon \,\omega _1 \,z\,\sin (2t -
\vartheta )
\\
 \dot {z} = \varepsilon \,\lambda - \varepsilon \,\gamma _2 \,z +
\varepsilon \,\omega _1 \,\alpha \;\sin \vartheta - \varepsilon
\,\omega _1
\,\alpha \,\sin (2t - \vartheta ) \\
 \dot {\vartheta } = \varepsilon \,\Delta - \varepsilon \left( {\omega _1
\,z / \alpha } \right)\;\cos \vartheta - \varepsilon \left(
{\omega _1 \,z /
\alpha } \right)\,\cos (2t - \vartheta ) \\
 \end{array}
\end{equation}

\noindent where a small frequency detuning is $\varepsilon \Delta
= 1 - \nu = \left( {\omega - \omega _0 } \right) / \omega $.

A method of averaging is applied to system (\ref{eq5})
\cite{6,10}. Following this method, we shall use evolutional
(drift) components $\bar {\alpha }\left( \tau \right),\,\bar
{z}\left( \tau \right),\,\bar {\vartheta }\left( \tau \right)$
($\tau = \varepsilon t$-slow time) in variables $\alpha \left( t
\right),\,\,z\left( t \right),\,\,\vartheta \left( t \right)$:

\begin{equation}
\label{eq6}
\begin{array}{l}
 \alpha \left( t \right) = \bar {\alpha }\left( \tau \right) + \varepsilon
u_1 \left( {\bar {\alpha },\bar {z},\bar {\vartheta },t} \right) + ... \\
 z\left( t \right) = \bar {z}\left( \tau \right) + \varepsilon \upsilon _1
\left( {\bar {\alpha },\bar {z},\bar {\vartheta },t} \right) + ... \\
 \vartheta \left( t \right) = \bar {\vartheta }\left( \tau \right) +
\varepsilon g_1 \left( {\bar {\alpha },\bar {z},\bar {\vartheta
},t}
\right)^ + ... \\
 \end{array}
\end{equation}

\noindent which satisfy to system of the evolutional (averaged)
equations of a form

\begin{equation}
\label{eq7}
\begin{array}{l}
 \dot {\bar {\alpha }} = \varepsilon { A}_1 \left( {\bar {\alpha },\bar
{z},\bar {\vartheta }} \right) + \varepsilon ^2{A}_2 \left( {\bar
{\alpha },\bar {z},\bar {\vartheta }} \right) + ... \\
 \dot {\bar {z}} = \varepsilon Z_1 \left( {\bar {\alpha },\bar {z},\bar
{\vartheta }} \right) + \varepsilon ^2Z_2 \left( {\bar {\alpha
},\bar
{z},\bar {\vartheta }} \right) + ... \\
 \dot {\bar {\vartheta }} = \varepsilon H_1 \left( {\bar {\alpha },\bar
{z},\bar {\vartheta }} \right) + \varepsilon ^2H_2 \left( {\bar
{\alpha
},\bar {z},\bar {\vartheta }} \right) + ... \\
 \end{array}
\end{equation}

With the use of averaging, we shall define the oscillative
corrections of the first order

\begin{equation}
\label{eq8}
\begin{array}{l}
 u_1 = - \left( {\omega _1 / 2} \right)\bar {z}\left( {\cos 2t\;\cos
\vartheta + \sin 2t\;\sin \vartheta } \right) \\
 \upsilon _1 = \left( {\omega _1 / 2} \right)\bar {\alpha }\left( {\cos
2t\;\cos \vartheta + \sin 2t\;\sin \vartheta } \right) \\
 g_1 = \left( {\omega _1 \bar {z} / 2\bar {\alpha }} \right)\left( {\sin
2t\;\cos \vartheta - \cos 2t\;\sin \vartheta } \right) \\
 \end{array}
\end{equation}

\noindent and equations for evolutional (drift) components $\bar
\alpha \left( \tau\right),\,\,\bar z\left( \tau\right),\,\,\bar
\vartheta \left( \tau\right)$ in the second order approximations

\begin{equation}
\label{eq9}
\begin{array}{l}
 \dot {\bar {\alpha }} = - \varepsilon \,\gamma _1 \,\bar {\alpha } -
\varepsilon \,\bar {z}\,\omega _1 \;\sin \bar {\vartheta } \\
 \dot {\bar {z}} = \varepsilon \,\lambda - \varepsilon \,\gamma _2 \,\bar
{z} + \varepsilon \,\omega _1 \,\bar {\alpha }\;\sin \bar {\vartheta } \\
 \dot {\bar {\vartheta }} = \varepsilon \left( {\Delta - \varepsilon \omega
_1^2 / 4} \right) - \varepsilon \left( {\omega _1 \,\bar {z} /
\bar {\alpha
}} \right)\;\cos \bar {\vartheta } \\
 \end{array}
\end{equation}

\subsection{Exact resonance condition -- nessesary and sufficient}

The third equation in system (\ref{eq9}) represents an analytical
form of a condition of strictly coherent interaction of two-level
system with a resonant field. This condition consists in constancy
in time of a difference of the current phases between a field and
atom, i.e. $\dot {\bar {\vartheta }} = 0$. At the same time this
condition $\dot {\bar {\vartheta }} = 0$ is a necessary and
sufficient condition for the exact resonance.

Let's pursue the brief analysis of the third equation. First term
is equality of detuning to zero ($\Delta = 0)$, i.e. equality of
frequencies $\omega = \omega _0 $ (which is sought in practice),
and only partially characterizes a resonance condition and is a
\textit{necessary} condition of the resonance. Third term $\delta
\omega \,^{(\ref{eq2})} = \omega _1^2 / 4$, the constant
correction, is the Bloch-Siegert shift. Second term $\delta \omega
\,^{(\ref{eq1})} = - \varepsilon \left( {\omega _1 \,\bar {z} /
\bar {\alpha }} \right)\;\cos \bar {\vartheta }$ - the variable
(drift) correction to resonance frequency in the first order. This
correction significant effects the resonance condition, i.e. a
\textit{two-level system can make a long-time drift of a resonant
frequency being in the state of zero detuning} $\Delta = 0$. Let's
write down a general form of equation for the resonance frequency
of a two-level system interacting with a weak harmonic field

\[
\omega = \omega _0 + \delta \omega \,^{(\ref{eq1})} + \delta
\omega ^{\,(\ref{eq2})} + ... \quad .
\]

\section{STEADINESS OF A STATIONARY RESONANT REGIME}

The consideration of a problem of existence and steadiness of the
stationary oscillation regimes in the system of equations
(\ref{eq9}) allows to make clear a possibility of practical
realization of sufficient conditions of resonance at $\dot {\bar
{\vartheta }} = 0$ (for performance of sufficient conditions it is
necessary to determine the stationary values of $\bar {\alpha }_s
,\,\,\bar {z}_s ,\,\,\bar {\vartheta }_s $ and accuracy of their
maintenance in time). The most important question is, in what
degree the equality $\bar {\vartheta }_s \left( {\tilde {\Delta }
= 0} \right) = \,\,\,\pm \pi / 2$ should be realized in practice.

\subsection{Existence of stationary resonant oscillation regimes}

Stationary regime, as is known \cite{10}, is characterized by an
invariance in time of the all output signal parameters -
amplitude, phase and frequency. Stationary values of variables,
designated as $\bar {\alpha }_s ,\,\,\bar {z}_s ,\,\,\bar
{\vartheta }_s $, are defined from system (\ref{eq9}). This system
of equations has the an unambiguous solution

\begin{equation}
\label{eq10}
\begin{array}{l}
 \bar {\alpha }_s = - \omega _1 \,\gamma _1 ^{ - 1}\,\bar {z}_s \;\sin \bar
{\vartheta }_s \\
 \bar {z}_s = \lambda / \left[ {\gamma _2 + \gamma \,\omega _1^2 \left(
{\gamma _1 ^2 + \tilde {\Delta }^2} \right)^{ - 1}} \right] \\
 \bar {\vartheta }_s = - arcctg\left( {\tilde {\Delta } / \gamma _1 }
\right) \\
 \end{array}
\end{equation}

From system (\ref{eq10}) we shall find the stationary values of
variables for a resonance:

\begin{equation}
\label{eq11}
\begin{array}{l}
 \bar {\vartheta }_s \left( {\tilde {\Delta } = 0} \right) = \,\,\,\pm \pi /
2\,\,\,\,\,\,\,\,\,\,\,\,\,\,\,\,\,\,\,\,\,\,\,\, -
difference\,\,of\,phase
\\
 \bar {z}_s \left( {\tilde {\Delta } = 0} \right) = \lambda / 2\gamma _2
\,\,\,\,\,\,\,\,\,\,\,\,\,\,\,\,\,\,\,\,\,\,\,\,\, -
\,difference\,\,of\,population \\
 \bar {\alpha }_s \left( {\tilde {\Delta } = 0} \right) = \lambda / 2\left(
{\gamma _1 \gamma _2 } \right)^{1 / 2}\,\,\,\,\,\,\,\,\,\, -
\,\,oscillation\,\,amplitude \\
 \end{array}
\end{equation}

The obtained result means that in system (\ref{eq9}) there is a
single stationary regime of sustained oscillations at the resonant
frequency.

\subsection{Steadiness of resonant oscillation regimes}

The stationary values for $\bar {\alpha }_s $ and $\bar {z}_s $
always exist because of the relaxation's terms ($\gamma _1
$,$\gamma _2 )$ in system of the equations (\ref{eq11}). The
similar conclusion for the difference of phases $\bar {\vartheta
}_s $ unequivocally cannot be made. There is a question arising:
with what accuracy must the equality $\bar {\vartheta }_s =
\,\,\,\pm \pi / 2$ take place to provide the conditions of
existing of the stationary regime (\ref{eq11}) and its steadiness
at which the slow drift of the frequency is excluded. The
quantitative estimation of an accuracy with which the equality
$\bar {\vartheta }_s \left( {\tilde {\Delta } = 0} \right) =
\,\,\,\pm \pi / 2$ would be executed the V.Volosov theorem about
stability of stationary resonant regimes of oscillations is given
\cite{10}:

In order to attain steadiness of a stationary resonance regime of
the "atom+field" system, initial conditions should be
simultaneously set for the three parameters, i.e. oscillation
amplitude, population difference, and the phase difference between
the field and the atom in the $\varepsilon $ -vicinity of their
stationary values. In so doing, the applied field frequency is
assumed to be constant, $\omega $= const ($\varepsilon $ is the
Rabi frequency-to-applied field frequency ratio).

When $\omega \ne $const, as is usually the case, the range of the
stationary regime steadiness decreases down to the $\varepsilon
^2$ -vicinity.

The impracticability of these conditions, in principle
($\varepsilon \sim 10^{ - 4}$ in radio spectroscopy and
$\varepsilon \sim 10^{ - 8}$ in optics) results in solely
non-steadiness operating regimes in spite of a great variety of
frequency stabilization systems in quantum standards. As a
consequence of the non-steadiness oscillation regime, a slow
resonance center frequency drift, unpredictable both in sign and
magnitude, appears in the system. This drift radically limits the
long-term stability of quantum metrology instruments.

The presence of non-steadiness even in a simple two-level system
should be taken into account in designing practical precise
quantum devices with high long-term stability.

\section{Summary}

The following results are achieved:

1). For the first time the existence of a variable component of
the first order is stated as regards the $H_1  \quad (E_1 )$
disturbance field amplitude in the resonance frequency of the two
level system interacting with a weak variable field.

2). As a result, necessary and sufficient conditions for the
accurate resonance are obtained being distinct from the commonly
accepted ones and topical for frequency standards and quantum
magnetometers. In particular, the most important sufficient
condition is a necessity of the coherence excitation with the
specified phase at the initial time instant.

3). When the sufficient conditions are not satisfied (a situation
which always takes place in practice), the oscillation regime with
a principally non-removable unsteadiness is performed in the
quantum system. It is expressed in drift and in found by the
author long-term variations of all output signal
parameters\textbf{.}

\begin{acknowledgments}
Thanks to Dr. J. Delporte, Dr. G. Mileti and my colleagues at RIRT
and SPU for their suggestions and encouragement. This work performed
at the St.-Petersburg State Polytechnic University, at financial
support with the INTAS - CNES under contract ¹ 03-53-5175.
\end{acknowledgments}

\end{document}